\documentclass[final,twocolumn,times]{elsarticle}
\usepackage{amssymb}

\usepackage{makecell}
 \usepackage{amsthm, amsmath}
 \usepackage{dsfont}
 \newtheorem{thm}{Theorem}
\newtheorem{lem}[thm]{Lemma}
\newtheorem{observation}[thm]{Observation}
\newdefinition{rmk}{Remark}
\newproof{pf}{Proof}
\newproof{pot}{Proof of Theorem \ref{thm2}}
\journal{Information Processing Letters}

\begin{document}
	
	\begin{frontmatter}
	
		
		
		\title{Geometric Separability using Orthogonal Objects}

	\author{V P Abidha \fnref{fn1} }
	\ead{abidha.vp@iiitb.org}
	\fntext[fn1]{First author is supported by the TCS Research scholar program}	
	%\address[2]{International Institute of Information Technology Bangalore, India}
%	\address[1][2]{International Institute of Information Technology Bangalore, India}
%\cortext[cor]{First author is supported by TCS Research scholar program}

		\author{Pradeesha Ashok}
		\ead{pradeesha@iiitb.ac.in}
		\address{International Institute of Information Technology Bangalore, India}
		

		\begin{abstract} 
Given a bichromatic point set $P=\textbf{R} \cup \textbf{B}$ of red and blue points,	a separator	is an object of a certain type that separates $\textbf{R}$ and $\textbf{B}$. We study the geometric separability problem when the separator is a) rectangular annulus of fixed orientation b) rectangular annulus of arbitrary orientation c) square annulus of fixed orientation d) orthogonal convex polygon. In this paper, we give polynomial time algorithms to construct separators of each of the above type that also optimizes a given parameter. 

\end{abstract}
		


		
	\end{frontmatter}

	

\section{Introduction}
		In many applications, data is represented using points in $\mathbb{R}^d$. In some of them, each data point belongs to a fixed number of categories or classes. In those cases, data from different classes can be represented by points of different colors. Thus the problems involving multicolored point sets are important in applications like data mining and machine learning.

In this paper, we study the separability problem for bichromatic point sets in the plane. Given a bichromatic point set $P=\textbf{R }\cup \textbf{B}$, where $\textbf{R}$ represents a set of red points and $\textbf{B}$ represents a set of blue points, the separability problem asks whether there exists a  geometric object of a particular type that separates $\textbf{R}$ and $\textbf{B}$. The separability problem is closely related to the classification problem in Machine Learning and has also found applications in areas like Data mining, Computer graphics, Pattern recognition, etc. An important question studied in separability is to decide whether a separator of a particular type exists for a given bichromatic point set. This question is studied for %For example, the linear separability question is to decide whether there exists a straight line such that the blue points and red points are in different half-planes defined by the line. 
%A generalization of this problem to higher dimensions is to check whether there exists a hyperplane that separates red and blue points. Both these problems can be solved in linear time~\cite{hyper}.
% The separability problem is also studied when the separator is 
  geometric objects like lines, hyperplane, circle, rectangle, convex polygon, etc.~\cite{hyper, Disk, rectangle, convex}. Another set of questions studied in separability is to find a separator that optimizes a given parameter like number of edges, area etc.~\cite{convex, OKM86}. 
  
 %A closely related problem that arises in the monochromatic setting is to find an annulus of minimum width that encloses a given point set. This question has been studied for square and rectangular annulus of fixed and arbitrary orientation~\cite{square, joydeep}. 

 Motivated by applications in urban scene reconstruction using LiDAR data, many researchers ~\cite{Van, rectangle,Urban}  considered the separability problem for various rectangular shapes. Rectilinear shapes are particularly important in urban scene reconstruction as many entities like roads and buidings tend to have orthogonal edges.  Sheikhi et al.~\cite{Farnaz}  studied the separability problem for a family of non-convex rectilinear objects called L-shapes and gave polynomial time algorithms. We consider a more general class of these objects, namely rectangular annulus. We also consider the separability problem for a class of objects called orthogonal convex objects, which is a restriction of classical convexity to an orthogonal setting.
 
% Applications that need to be separated will benefit from effective results to many problems. For example, city planning, where red points represent buildings and monuments, the goal is to build a jogging path and a park that surrounded by  the monuments. Another application is the facility location problem, in which we have different types of facilities and our objective is to identify the location of rectangular shapes. 
 
% suppose tissue containing tumor, where both tumor cells and healthy cells that defined by their coordinates. Our aim is to separate the tumor cells from the healthy cells in radiation therapy or surgical removal. .% Also, we can use bi chromatic problems in finding the maximum material boundary in manufacturing engineering and 

\noindent\textbf{Problems Studied and Related Work:}

%------------Rectangle annulus--------------------------

Let $P=\textbf{R} \cup \textbf{B}$ be a point set in $\mathbb{R}^2$ where $\textbf{R}$ is a set of red points and $\textbf{B}$ is a set of blue points and $|\textbf{R}|+|\textbf{B}|=n$, throughout the paper. For simplicity of explanation, we assume no two points in $P$ have the same $x$ or $y$ coordinate. 

\noindent \textbf{The separability problem} is to find an object $C$ such that all the blue points lie inside or on the boundary of $C$ and all the red points lie outside or on the boundary of $C$. We study the separability problem for annuli. In general, an annulus refers to the region enclosed between two objects of the same type. In this paper, we study the separability problem for Rectangular and Square annuli of fixed and arbitrary \emph{orientation}. We define the \emph{orientation} of a line as the angle it makes with the $X$-axis. A rectangle/square (and a rectangular/square annulus) has orientation $\theta$ if the line containing each of its edges has orientation either $\theta$ or $\theta+\pi/2$. We will now define these objects.

\noindent\textbf{Rectangular Annulus}: A rectangular annulus is a closed region between two rectangles $D$ and $D'$  of same orientation such that $D$ is contained in $D'$. %and each edge of $R$ is parallel to the edge of $R'$ on the same side.
For a rectangular annulus $A$, we shall call the larger rectangle and smaller rectangle that define $A$ as its outer rectangle and inner rectangle respectively. 
%that defines $R$ as its outer rectangle and the smaller rectangle that defines $R$ as its inner rectangle. 
The \emph{width} of a rectangular annulus on a given side is the perpendicular distance between the corresponding edges of the outer and inner rectangles.  If the width is uniform on all the four sides of an annulus, we call it a uniform width rectangular annulus. Note that the center of inner and outer rectangles are the same in a uniform width rectangular annulus (The converse is not always true). When the width is not uniform on all the four sides, the width of the annulus is defined as the maximum width among the four sides.

\noindent\textbf{Square Annulus}: A square annulus is a closed region between two squares $S$ and $S'$ of same orientation such that $S$ is contained in $S'$. The definition of the \emph{width} of the square annulus is similar to that of the rectangular annulus. 
If the center (point of intersection of the diagonals) of the outer and inner squares are the same, or equivalently the width is uniform on all the four sides, then we call it a  concentric square annulus.

For a given bi-chromatic point set, we aim to find separating rectangular and square annuli that minimize the width. The Separability problem using annulus can be considered as a natural extension of a closely related problem that arises in the monochromatic setting, namely, to find an annulus of minimum width that encloses a given point set. This question has found applications in many areas including facility location~\cite{FacilityLocation}, tolerancing metrology, urban planning~\cite{city}, rectangle fitting~\cite{best} etc. and has been studied for square and rectangular annulus of fixed and arbitrary orientation~\cite{square, joydeep}. Our results extend many of these results for the case where there are \emph{unwanted} entities that need to be separated. Consider an example of surveillance of military establishments along a path of orthogonal visibility that avoids civilian establishments or constructing a rectangular path in a tourist spot that contains all monuments but avoids all trees.

Many techniques used for solving the monochromatic problems can be extended for solving the minimum width separating annulus problem. However, the presence of red points poses certain challenges in the separability problem. For instance, it is easy to extend a nonuniform width annulus to a uniform width annulus in the monochromatic setting whereas the two cases are very different in the bichromatic setting. The algorithms presented in this paper makes non-trivial modifications to the existing techniques to solve the bichromatic separability problem. Moreover, the running times of the algorithms presented are the same as that of the best known running times of the corresponding monochromatic variants.
%To the best of our knowledge, the separability problem is not studied for annuli. In related objects,~\cite{Farnaz} gives polynomial time algorithms when separating objects are L-shapes, which is a special case of rectangular annulus separabilty. This problem was motivated by applications in Urban area reconstruction.

% \noindent\textbf{Orthogonal Convex Hull}: The orthogonal convex hull of a point set $P$ is defined as the smallest area orthogonal convex polygon that contains $P$. It follows from the definition of an orthogonal convex hull that its convex vertices are points from $P$. The orthogonal convex hull of a point set with $n$ points can be computed in $O(n \log{n})$ time~\cite{Karlsson}.

We also consider the separability problem for orthogonal convex objects.

%Motivated by applications in urban scene reconstruction using LiDAR data, many researchers ~\cite{Van, Urban}  considered the separability problem for various rectangular shapes. Rectilinear shapes are particularly important in urban scene reconstruction as many entities like roads and buidings tend to have orthogonal edges.% In addition to the papers mentioned above, Sheikhi et al.~\cite{Farnaz}  studied the separability problem for a family of non-convex rectilinear objects called L-shapes and gave polynomial time algorithms. We consider a more general class of these objects, namely rectangular annulus. We also consider the separability problem for a class of objects called orthogonal convex objects, which is a restriction of classical convexity to an orthogonal setting.

\noindent\textbf{Orthogonal Convex Polygon}: A polygon $C$ is said to be orthogonal convex if all the edges of $C$ are either horizontal or vertical and the intersection of any vertical or horizontal line with $C$ is either null, a point or a continuous line segment~\cite{Burkay}. An orthogonal convex polygon has alternate vertical and horizontal edges and alternate convex and reflex vertices (except for the vertices with highest and lowest $x$ and $y$ coordinates).

A polynomial time algorithm to find a separating convex polygon with the minimum number of edges is given in~\cite{convex}. We extend this result to orthogonal convexity. Note that minimizing the number of edges is equivalent to minimizing the number of bends in the enclosing structure which is useful in areas like circuit designing.

	\label{}
	%\section{preliminaries}
	%\vspace{5pt}
\noindent A summary of the results of this paper are given in Table~\ref{tab}. All the results for the separability problem for minimum width annulus can be extended to the separability problem for maximum width annulus.

%\begin{center}
	\begin{table}
	\begin{tabular}{|l|c|} 
		\hline
		Object&  Complexity\\
		\hline
		 Rectangular annulus(NU) (F)& $O(n)$\\
		 Rectangular annulus (U) (F) & $O(n\log n)$\\
	   	 Rectangular annulus(NU) (A) & $O(n^2 \log n)$\\
	     Square Annulus(U)(F) &$O(n\log^2 n)$\\
		Orthogonal convex polygon(F)& $O(n\log n)$\\
		\hline
	\end{tabular} 
	\caption{\label{tab} Summary of the results (NU and U represent nonuniform and uniform width resp. and  F and A represent fixed and arbitrary orientation resp.)}
	\end{table}
% \end{center}

\noindent\textbf{Notations}: For any point $p \in P $, let $x(p)$ and $y(p)$ respectively denote the $x$-coordinate and $y$-coordinate values of $p$ . We further use the same notation for vertical and horizontal lines. For a rectangle $D$, let $l(D)$, $b(D)$, $r(D)$ and $t(D)$ denote the left, bottom, right and top edges of $D$ respectively. For a rectangular (square) annulus $A$, we refer to the left boundary of $A$ i.e., the union of left edges of the outer and inner rectangles (squares) of $A$, as the left side of $A$ (similar for
other sides). For a point set $P$, $CH(P)$ denotes the convex hull of $P$.

	\section{Rectangular Annulus Separability in Fixed Orientation}
	\label{RAS}

%In this section, we consider the Rectangular Annulus Separability problem. 
%For a given bi-chromatic set P, we give an algorithm to find a non-uniform and uniform width separating axis-parallel rectangular annulus $A$ of minimum width.
%We further give an algorithm to find a uniform width separating rectangular annulus $A$ of minimum width.

%%%
%\begin{figure}[ht]
	%	\begin{minipage}[b]{0.50\linewidth}
%	\centering
%	\includegraphics{nonconcentric1.pdf}
%	\scalebox{0.55}{\input{nonconcentric1.pdf_t}}
%	\caption{Non-uniform width rectangular annulus of blue points}
%	\label{fig:figure3}

%\end{figure}

%%%

%----------------

%\begin{center}
 %\begin{figure}[h]
%\scalebox{0.4}{\input{nonconcentric1.pdf_t}}
%\caption{Non-uniform width rectangular annulus of blue points}
%\end{figure}
%\end{center}

Let $B_{out}$ be the smallest axis-parallel rectangle that contains all points of $\textbf{B}$. Let $\textbf{R}^\prime{}$ be the set of red points inside $B_{out}$ and $R_{in}$ be the smallest axis-parallel rectangle that contains $\textbf{R}^\prime{}$. If $R_{in}$ contains blue points inside, then a separating rectangular annulus does not exist. If $\textbf{R}^\prime{}$ is empty, then the problem can be solved by the algorithm that computes the minimum width enclosing rectangular annulus given in~\cite{joydeep}.  %Mukherjee et al.~\cite{joydeep} gave an $O(n)$ algorithm that computes minimum width enclosing axis-parallel rectangular annulus

%The following lemma is a special case of section~\ref{arb}.

%If there is only one red point in $\textbf{R}^\prime{}$ then consider it as a rectangle such that all sides contain the same red point. 

%If $R'$ is empty, then the problem can be solved by the algorithm that computes the minimum width enclosing rectangular annulus given in \cite{joydeep}.

%$B_{out}$ contains the leftmost, bottommost, rightmost and topmost points of $\textbf{B}$ in the boundary.  Similarly $R_{in}$ contains the leftmost, bottommost, rightmost and topmost points of $\textbf{R}^\prime{}$ in the boundary. Thus $B_{out}$ and $R_{in}$ can be constructed in $O(n)$ time. All other degenerate cases can be done in $O(n)$ time.
 
 %---------append
 
\begin{lem}
\label{nonlemma1}
A separating axis-parallel  (non-uniform width) rectangular annulus  of minimum width (if exists) can be constructed in $O(n)$ time. 
\end{lem}

\begin{pf}

 Let $T_A$ (resp. $B_A$) be the set of all blue points $p_i$ inside $B_{out} $ such that $x(l(R_{in})) \leq x(p_i)\leq$  $x(r(R_{in}))$ and $y(p_i)\geq y(t(R_{in}))$ (resp. $ y(p_i) \leq y(b(R_{in}))$). Let $R_{A}$ (resp. $L_A$) be the set of all blue points $p_i$ inside $B_{out} $ such that $y(b(R_{in})) \leq y(p_i) \leq y(t(R_{in}))$ and $ x(p_i) \geq x(r(R_{in}))$ (resp. $ x(p_i) \leq x(l(R_{in}))$).

Define the set $T(B_{out} )$ as the union of $T_A$ and two sets $T_1$ and $T_2$, where $T_1$(resp. $T_2$) contains the set of all points $p_i \in \textbf{B} \backslash (B_A \cup R_A \cup L_A)$ with $ y(p_i)>y(t(R_{in}))$ and  $x(p_i)>x(r(R_{in}))$ (resp. $x(p_i) < x(l(R_{in}))$) such that the perpendicular distance between $p_i$ and $t(B_{out})$ is not larger than the perpendicular distance between $p_i$ and $r(B_{out})$ (resp. $l(B_{out})$). Similarly define the sets $B(B_{out} )$, $R(B_{out} )$, and $L(B_{out})$ corresponding to the bottom, right and left edges of $B_{out}$.  Let $d(p,l)$ be the perpendicular distance between $p$ and $l$, where $p$ is a point and $l$ is a vertical or horizontal line. 
%Therefore $T(B_{out})= T_{A} \cup \{ p_i| d(p_i,t(B_{out})) \leq \min \{d(p_i, r(B_{out})), d(p_i, l(B_{out}))\}$ where $p_i \in \textbf{B} \backslash (B_A \cup R_A \cup L_A)\}$ and $ y(p_i)>y(b(R_{in}))$. 
Let $p_t'$ be the element of $T(B_{out})$ such that $d(p_t', t'(B_{out}))= \max \{ d(p_i, t'(B_{out}))|p_i \in T(B_{out})\}$, where $t'(B_{out})$ is the line containing the line segment $t(B_{out})$. Similarly define $p_b'$, $p_r'$ and $p_l'$ from $B(B_{out})$, $R(B_{out})$ and $L(B_{out})$ respectively. Construct $B_{in}$ as the rectangle that contains $p_t'$, $p_b'$, $p_r'$ and $p_l'$ on its top, bottom, right and left edges respectively. Now the annulus $A^\prime{}$, defined by $B_{in}$ and $B_{out}$ is a separating rectangular annulus. Moreover, by construction, any other rectangular annulus with width less than that of $A^\prime{}$ cannot contain all points from \{$p_t'$, $p_b'$, $p_r',$ $p_l'$ \}, therefore $A^\prime{}$ is a separating rectangular annulus with minimum width. 
%as any other annulus of smaller width will leave out at least one of the above defined points. Since 
$B_{in}$, $R_{in}$ and $B_{out}$ can be constructed in $O(n)$ time.
\qed
\end{pf}

%\begin{figure}[h]
%	\centering
%	\scalebox{.30}{\input{blue_annulus_without_Rout.pdf_t}}
%	\caption{Separating uniform width axis parallel rectangular annulus with anchor points. (the dotted lines show the annulus after moving)}
%	\label{fig:figure4}
%	\end{figure}

%We now consider the Rectangular Annulus Separability problem for uniform width axis-parallel rectangular annulus.
% Finding a minimum width enclosing rectangular annulus is trivial one, a non-uniform width enclosing rectangular annulus is computed as we can transform a non-uniform width enclosing rectangular annulus to a uniform width one, by moving some set of the edges the outer rectangular. However, thus is not possible in separability due to the presence of red points outside.

%Let $A$ be a uniform width separating rectangular annulus (if it exists) with minimum width.

%Let $A_{out}$ and $A_{in}$ respectively be the outer and inner rectangles of $A$. Note that $A_{out}$ contains all the blue points inside and  $A_{in}$ contains all the points of $R^\prime{}$ and no blue point inside.
\noindent \emph{Uniform width axis-parallel rectangular annulus : }
Let $\mathcal{B}_{out}$ be the set of (at most) four blue points that lie on $t(B_{out})$, $b(B_{out})$, $r(B_{out})$ and $l(B_{out})$. Similarly define the set $\mathcal{R}_{in}$ for $R_{in}$.
	Let $F_p = \mathcal{B}_{out}\cup \mathcal{R}_{in}$. We call the points in $F_p$ as \textit{fixed points}. Note that an input point lying on a corner can be considered to be lying on two sides. For example, an input point on the top-left corner of a rectangle is considered as a point on the top edge as well as a point on the left edge.

	\begin{thm}
		\label{unitheorem1}
		A separating uniform width axis-parallel rectangular annulus of minimum possible width (if exists) can be constructed in $O(n \log n)$ time.
	\end{thm}
	
	\begin{pf}
We begin by observing that there exists a separating uniform width  axis-parallel rectangular annulus of minimum possible width $A$ that contains $5$ points from $\textbf{R} \cup \textbf{B}$ on its boundary. We call these points as the \textit{anchor points} of $A$. 	
		%Let $A$ be a  separating uniform width  axis-parallel rectangular annulus of minimum possible width. 
Let $A_{out}$ and $A_{in}$ represent the outer and inner rectangles of $A$ respectively. Then both $A_{out}$ and $A_{in}$ contain at least one blue point on the boundary. Otherwise, we can expand $A_{in}$ (shrink $A_{out}$) to get a rectangular annulus of smaller width. Consider any side of $A$. If this side does not contain a fixed point on its boundary, then we can move this side parallel to itself, towards the centre of the annulus till it touches a point from $\mathcal{B}_{out}$ or $\mathcal{R}_{in}$. It is easy to see that the annulus still contains the same set of points inside. Therefore, we can assume that $A$ contains a fixed point each on its four sides and a blue point on the boundary of $A_{in}$, which together form the $5$ anchor points of $A$. 
			The set of four anchor points which are also fixed points can be determined in $O({8\choose 4})$ time.  The fifth anchor point belongs to the set $\textbf{B} \backslash F_{p}$, since $A_{in}$ contains a blue point.  This point can be determined in $O(n)$ time. Thus the set of five anchor points can be correctly guessed in $O({8\choose 4}n)$ time. For a given set of five points, the uniform width rectangular annulus that contains them on the boundary can be constructed in constant time.  By using a data structure like range tree, it can be checked whether a given rectangular annulus is a separating rectangular annulus by a constant number of orthogonal range searching queries (We need to check whether $A_{in}$ contains any blue point and $A_{out}$ contains any red point from $\textbf{R }\setminus\textbf{R}'$). A range tree can be constructed in $O(n \log n )$ time and each range searching query can be completed in $O(\log n)$ time~\cite{Berg}. Thus all valid separating rectangular annuli can be constructed in $O(n \log n)$ time and we can return the one with the minimum width.
	\qed
		\end{pf}

	\section{Rectangular Annulus Separability in Arbitrary Orientation }
	\label{arb}
	
In this section, we give an algorithm based on the rotating calipers method~\cite{rotatingcalipers} to find a separating (non-uniform width) rectangular annulus of arbitrary orientation of minimum width.

Let $MER(P)^{\theta} $ be the minimum enclosing rectangle of a point set $P$ in orientation $\theta$. $MER(P)$ in any orientation contain points from $CH(P)$ on its four edges, hereafter referred as the \emph{anchor points} of $MER(P)$.

\begin{lem}
	\label{semicircle}
	Let $P, Q$ be two sets of points such that points in $Q$ lie outside $CH(P)$. Then all values of $\theta \in [0,\frac{\pi}{2}]$ where a given point $q \in Q$ lies inside  $MER(P)^{\theta} $ form a continuous interval. 
	%Moreover, for all values of $\theta$ in this interval, $MER(P)^{\theta} $ contains the same set of points from the convex hull of $P$ on its edges.
\end{lem}

\begin{pf}
	Let  $t$, $r$, $b$, $l \in CH(P)$ respectively be the points that are the top, right, bottom and left anchor points of $MER(P)^{\theta} $. The set of orientations $\theta$ where  $t$, $r$, $b$, $l $ are the anchor points of $MER(P)^{\theta} $ form a continuous interval, say $[\alpha_1,\alpha_2]$, defined by the orientations of the edges of $CH(P)$~\cite{joydeep}. 
	% For a point $q$ outside $CH(P)$ to lie in $MER(P)$ in a given orientation, it should lie in the wedges given by two adjacent edges of $CH(\textbf{B})$ with the common point as a anchor point. 
	Consider two of these points which lie on adjacent sides of an $MER$, say $t$ and $r$. Then the top right corner of $MER(P)^{\theta} $ will lie on a semicircle $M$ with  $t$ and $r$ as diameter points. We can consider three more similar semicircles such that any point of $Q$ that lies inside $MER(P)^{\theta} $ will also lie in the interior of one of them. Let $q\in Q$ be a point that lies in the interior of $M$. Consider the point $q_1$(resp. $q_2$) where the line connecting $q$ and $r$(resp. $t$) intersects the boundary of $M$. 
	%Similarly let $q_2$ be the point where  the line connecting $q$ and $t$ intersect the boundary of $M$. 
	Then, $q$ lies inside any $MER(P)^{\theta}$ whose top-right corner lies between $q_1$ and $q_2$ on the boundary of $M$. This happens when  $\theta \in [\gamma_1, \gamma_2] \cap [\alpha_1,\alpha_2]$, where $\gamma_1, \gamma_2$  are the orientations of the lines $tq_1$ and  $tq_2$ respectively. 
	
	For $i \in \{1,2,3\}$, let $t_i, r_i$ respectively be the top and right anchor points of $MER(P)^{\theta} $ when $\theta$ belongs to an interval $[\alpha_1^i,\alpha_2^i]$. Let the intervals be such that $\alpha_1^1 < \alpha_1^2 < \alpha_1^3$ and for $i \in \{1,2\}$, either $t_i \ne t_{i+1}$ or $r_i \ne r_{i+1}$ (or both). It is easy to see that if $q$ lies inside $MER(P)^{\theta_1}$ and $MER(P)^{\theta_3}$ for some $\theta_1 \in  [\alpha_1^1,\alpha_2^1]$ and some  $\theta_3 \in  [\alpha_1^3,\alpha_2^3]$ then $q$ lies inside $MER(P)^{\theta_2}$ for all  $\theta_2 \in  [\alpha_1^2,\alpha_2^2]$. Therefore, all values of $\theta$, where $q$ lies inside $MER(P)^{\theta}$ form a continuous interval in $[0,\frac{\pi}{2}]$. To define this interval, let $t_i, r_i$ be the top and right anchor points of $MER(P)^{\theta} $ when $\theta \in [\alpha_1^i,\alpha_2^i]$ and $\alpha_1^i < \alpha_1^{i+1}$ for $1\leq i < k$. Let $M_i$ be the semicircle with $t_i$ and $r_i$ as end points. If $q$ lies inside all $M_i$'s, for $1 \leq i \leq k$, then $q$ lies inside $MER(P)^{\theta}$ for $\theta \in [\gamma_1, \gamma_2]$, where $\gamma_1$ is the orientation of the line connecting $q$ and $t_1$ and $\gamma_2$ is the orientation of the line connecting $q^\prime$ and $t_k$, where $q^\prime$ is the intersection point of the line connecting $q$ and $ r_k$ on $M_k$.
\qed

%	Let $t_x,r_x$, $t_y,r_y$ and $t_z,r_z$ in $CH(P)$ be the points which lie on the top and right edges on $MER(P)^{\theta} $ in the intervals $[\alpha_x, \alpha_{x'}]$, $[\alpha_y, \alpha_{y'}]$, and $[\alpha_z, \alpha_{z'}]$ respectively.
%	
%	If $q \in Q$ lies inside $MER(P)^{\theta_{1}} $ and $MER(P)^{\theta_{2}} $, where $\theta_{1} \in [\alpha_x, \alpha_{x'}]$  and $\theta_{2} \in [\alpha_z, \alpha_{z'}]$, then for all values of $\alpha_y$ and $ \alpha_{y'}$ , such that $\alpha_x < \alpha_y <  \alpha_{z}$ and $\alpha_{x'} < \alpha_{y'} < \alpha_{z'}$. $q$ lies inside $MER(P)^{\theta} $, for all values of $\theta \in [\alpha_y, \alpha_{y'}]$.
%	i.e, for all values of $\theta$, where  $q$ lies inside $MER(P)^{\theta}$ form a continuous intervals in $[0,\frac{\pi}{2}]$.
%	
%	
%	Let $L(p,q)$ be the line passes through the point $p$ and $q$.
%	Let $t_1, t_2, . . . , t_k$ be the top points in the non decreasing order of the orientation of $L(q,t_{i})$, where $1\leq i\leq k$. i.e, $L(q,t_{1}) \leq L(q,t_{2}) \leq . . . , L(q,t_{k})$. Similarly the right points are in the order $r_1, r_2, . . . r_k$  and $L(q,r_{1}) \leq L(q,r_{2}) \leq . . . ,  L(q,r_{k})$ respectively.
%	
%	Union of all invalid intervals created by $q$ in $[0,\frac{\pi}{2}]$ is same as the invalid interval $[\gamma_1, \gamma_2]$, where $\gamma_1$ is the orientation of $L(q, t_1)$  and $\gamma_2$ is the orientation of $L(q^\prime, t_k)$, where $q^\prime$ is the intersection point of $L(q, r_k)$ on the semicircle defined by $r_k$ and $t_k$.
%	
%	
%	

\end{pf}
\noindent\textit{Computing All Orientations where a Separating Rectangular Annulus exists : }
Let $\textbf{R}^\prime$ be the set of red points inside  $CH(\textbf{B})$. If $CH(\textbf{R}^\prime)$ contains a blue point inside,  no separating rectangular annulus exists. 

Let $A^{\theta}$ be  a separating rectangular annulus of minimum width in a fixed orientation $\theta$. Then $A_{out}^{\theta}$, the outer rectangle of $A^{\theta}$, is $MER(\textbf{B})^{\theta}$. Similar to the algorithms given in \cite{joydeep}, we subdivide the interval $[0, \frac{\pi}{2}]$ into \emph{primary intervals}  such that for each orientation $\theta$ in a primary interval the set of anchor points of $MER(\textbf{B})^{\theta}$
%the minimum enclosing rectangle in that orientation 
is the same. 
The number of primary intervals is $O(k)$ where $k$ is the number of points on $CH(\textbf{B})$.

Let $B_{out}^{\theta}$ be $MER(\textbf{B})^{\theta}$ and $t_{b}^{\theta}$, $r_{b}^{\theta}$, $b_{b}^{\theta}$ and $l_{b}^{\theta}$ be the blue points on the boundary of $B_{out}^{\theta}$, for $\theta \in [0,\frac{\pi}{2}]$. We define $\mathcal{B}_{out}^{\theta} = \{t_{b}^{\theta}$,$r_{b}^{\theta}$,$b_{b}^{\theta}, l_{b}^{\theta}\}$.
Let $\textbf{R}_1^{\theta}$ be the set of red points inside $A_{out}^{\theta}$ for a given $\theta$. Assume $\textbf{R}_1^{\theta}$ is non-empty (The case where $\textbf{R}_1^{\theta}$ is empty is discussed later). The set $\textbf{R}_1^{\theta}$ may vary in a primary interval based on the value of $\theta$. This happens as some red points that lie outside $CH(\textbf{B})$ may lie  inside $MER(\textbf{B})^{\theta}$ for some values of $\theta$. For such a red point $r$, the values of $\theta $ for which $r$ lies in $MER(\textbf{B})^{\theta}$ forms a continuous subinterval of $[0,\frac{\pi}{2}]$, by Lemma~\ref{semicircle}.

In the next step, we further subdivide the primary intervals based on the four points in $\textbf{R}_1^{\theta}$ with the smallest perpendicular distance to each side of $A_{out}^{\theta}$. For $p,q \in \mathbb{R}^2$, let $d(p,q,\theta)$ denote the perpendicular distance between $p$ and a straight line of orientation $\theta$ that passes through $q$. For a fixed $p$ and $q$, this function is a continuous sinusoidal function. Based on this function, we define $d'(r,., \theta)$, a distance function for all red points ($r \in R$) as follows:

\begin{equation} 
d'(r,t_b^{\theta},\theta)= \left\{
\begin{array}{cc}
\infty & if  r \notin \boldsymbol{\textrm{R}_{1}^{\theta}}\\  
d(r,t_b^{\theta},\theta), & otherwise 
\end{array} 
\right.
\end{equation}

Note that $d'(.,.,.)$ is also sinusoidal and two functions intersect at most once (Lemma 1, in~\cite{square}). Now if we plot the function $ d'(r,t_b^{\theta},\theta)$ for all $r \in R$, the lower envelope gives the nearest point to $t(A_{out}^\theta)$ for all $\theta$. Similarly, repeat for other sides of $A_{out}^{\theta}$. Thus for every $\theta$, we get a set of four points $\mathcal{R}_{in}^{\theta} =\{t_r^{\theta}, l_r^{\theta}, b_r^{\theta}, r_r^{\theta}\}$ that defines the smallest rectangle of orientation $\theta$, $R_{in}^{\theta}$, that contains all points in $\textbf{R}_{1}^{\theta}$. We subdivide the primary intervals to \emph{secondary intervals}, such that for every angle $\theta$ in a secondary interval, the set $\mathcal{B}_{out}^{\theta} \cup \mathcal{R}_{in}^{\theta}$ is the same. 

For a given $\theta$, if $R_{in}^{\theta}$ contains any blue point inside, then there is no separating rectangular annulus in orientation $\theta$. In $O(n)$ time we can verify that whether there exists a blue point inside $CH(\mathcal{R}_{in}^{\theta})$ or not. There may be some blue points lying outside $CH(\mathcal{R}_{in}^{\theta})$ and inside $R_{in}^{\theta}$. Such points will also lie in the interior of a semicircle with two anchor points from $\mathcal{R}_{in}^{\theta}$ as diameter points. By lemma~\ref{semicircle}, for every blue point $b$ outside $CH(\mathcal{R}_{in}^{\theta})$, we can compute the interval where $b$ lies inside $R_{in}^{\theta}$ within a secondary interval.
Such an interval is now an \emph{invalid interval} in that if $\theta$ belongs to that interval then there is no separating rectangular annulus of orientation $\theta$. We remove the invalid interval corresponding to every blue point from the set of secondary intervals to get a set $\mathcal{I}$ of \emph{valid} sub-intervals. Now the following result is easy to see.

 \begin{lem}
 	\label{allOrientations}
 For all the orientations $\theta$ that belong to a valid sub-interval in $\mathcal{I}$, there exists a rectangular annulus of orientation $\theta$ which is a separating rectangular annulus for $P$.
 \end{lem}

%\noindent\textbf{Proof of Lemma \ref{allOrientations}}

%\begin{pf}
%Let $\theta$ belong to a sub-interval in $\mathcal{I}$. Then there exist eight points in $\mathcal{B}_{out}^{\theta} \cup \mathcal{R}_{in}^{\theta}$ that will define a rectangular annulus, $N^{\theta}$. We will show that $N^{\theta}$ is a separating rectangular annulus for $P$. The outer rectangle of $N^{\theta}$ is a minimum enclosing rectangle of $\textbf{B}$ and contains all blue points inside. Also, the inner rectangle contains all red points of $\textbf{R}_i^{\theta}$ inside since we have selected the points with the smallest perpendicular distance to each side of the outer rectangle. Also for each blue point, we have excluded the angles where it comes inside the inner rectangle. 
%\qed
%\end{pf}

%\begin{pf}
%Let $\theta$ belong to a sub-interval in $\mathcal{I}$. Then there exist eight points in $\mathcal{B}_{out}^{\theta} \cup \mathcal{R}_{in}^{\theta}$ that will define a rectangular annulus, $N^{\theta}$. We will show that $N^{\theta}$ is a separating rectangular annulus for $P$. The outer rectangle of $N^{\theta}$ is a minimum enclosing rectangle of $B$ and contains all blue points inside. Also, the inner rectangle contains all red points of $\textbf{R}_1^{\theta}$ inside since we have selected the points with the smallest perpendicular distance to each side of the outer rectangle. Also for each blue point, we have excluded the angles where it comes inside the inner rectangle. 
%
%\qed
%\end{pf}

\noindent\emph{Constructing a Rectangular Annulus of Minimum Width :}
Now, we will construct $A^{\theta}$, the separating annulus of minimum width,  for each orientation $\theta$. Note that we know the outer rectangle of $A^{\theta}$ is same as $B_{out}^{\theta}$. 
For each side of $B_{out}^{\theta}$, we will define a set of blue points that will lie closer to that side in a minimum width separating rectangular annulus. % Consider the side $t(B_{out}^{\theta})$ that contains the point $t_b^{\theta}$. Let $T_A$ intuitively be the set of blue points that will lie closer to $t(B_{out}^{\theta})$ in a minimum width separating rectangular annulus.
Let $T_1$ be the set of points that lie inside a rectangle defined by the lines containing edges $t(R_{in}^{\theta})$, $l(R_{in}^{\theta})$, $r(R_{in}^{\theta})$ and $t(B_{out}^{\theta})$. We will refer to these points as \emph{special top} points. 
%Clearly, the special top points belong to the set $T_A$.
 Similarly, we can define the sets $B_1, L_1$ and $R_1$ as special bottom points, special left points and special right points.
 % Of the remaining blue points, 
 Let $T_2$ be the set of all blue points that lie inside the top-right rectangle (resp. top-left rectangle) defined by the lines containing edges $t(R_{in}^{\theta})$, $r(R_{in}^{\theta})$, $r(B_{out}^{\theta})$ and $t(B_{out}^{\theta})$ (resp. $t(R_{in}^{\theta})$, $l(R_{in}^{\theta})$, $l(B_{out}^{\theta})$ and $t(B_{out}^{\theta})$) whose perpendicular distance to $t(B_{out}^{\theta})$ is smaller than the perpendicular distance to $r(B_{out}^{\theta})$ (resp. $l(B_{out}^{\theta})$) (Similarly define the sets $B_2, L_2$ and $R_2$). 
 
 Let $T_A=\{T_1 \cup T_2\}$ (Similarly define the sets $B_A, L_A$ and $R_A$). 
 %, intuitively be the set of blue points that will lie closer to $t(B_{out}^{\theta})$ in a minimum width separating rectangular annulus
 Thus for any valid orientation $\theta$, the sets $T_A,B_A, L_A, R_A$ gives a partition of all blue points. For a given point $b$, we can compute the angular interval where it is a special point in constant time. When a point $p$ is not a special point, it belongs to the set $T_A$ if $min(d(p,t_b^{\theta}, \theta), d(p,r_b^{\theta}, \theta)) = d(p,t_b^{\theta},\theta)$ for $p$ lying inside the rectangle defined by the lines containing edges $t(R_{in}^{\theta})$, $r(R_{in}^{\theta})$, $r(B_{out}^{\theta})$ and $t(B_{out}^{\theta})$ and if $min(d(p,t_b^{\theta}, \theta), d(p,l_b^{\theta}, \theta)) = d(p,t_b^{\theta},\theta)$, for $p$ lying inside the rectangle defined by the lines containing edges $t(R_{in}^{\theta})$, $l(R_{in}^{\theta})$, $l(B_{out}^{\theta})$ and $t(B_{out}^{\theta})$. Now consider the following function,

\begin{equation}
f_1(p,t,\theta)=\left\{
\begin{array}{cc} 
d(p,t_b^{\theta},\theta) & if  p \in T_A\\
0 & otherwise
\end{array} 
\right.
\end{equation}

The upper envelope of the set of functions $f_1(p,t,\theta)$ for all $p \in \textbf{B}$ gives the width of the top side of a separating annulus of minimum width for each $\theta$. By similar methods, we can compute the width of all sides of a separating annulus of minimum width. The value of $\theta$ that minimizes the width can be returned.

\begin{thm}
For a bichromatic point set $P$, a minimum width separating rectangular annulus of arbitrary orientation can be computed in $O(n^2 \log n)$ time.
\end{thm}

\begin{pf}
The algorithm computes the set of primary intervals in $O(n\log n)$ time~\cite{joydeep}. In order to compute the set of secondary intervals, we compute the function  $d'(r,*,*)$ for all $r \in \textbf{R}$ in $O(n)$ time and compute the lower envelope of these functions. This can be done in $O(n\log n)$ time since the graphs of these functions have bounded intersection~\cite{square}. 
Any secondary interval can be labeled with exactly one of the following events thus showing that there are $O(n)$ secondary intervals.\\
1. Addition/Removal of a point from the set $\mathcal{B}_{out}$ .\\
2.  Addition/Removal of a red point that lies inside $CH(\textbf{B})$ to the set $\mathcal{R}_{in}^{\theta}$.\\
3. Addition/Removal of a red point that lies outside $CH(\textbf{B})$ to the set $\mathcal{R}_{in}^{\theta}$.\\
Computing all valid intervals includes finding the invalid subintervals corresponding to each blue point and merging all those intervals, which can be done in $O(n\log n)$ time. There are $O(n)$ invalid intervals and therefore, $O(n)$ valid intervals. Now, for a valid interval $I \in \mathcal{I}$, we need to calculate the upper envelope of all the functions $f_1(p,*,\theta)$ for all $p \in \textbf{B}$. Since the distance function is a sinusoidal function, this can be done in $O(n\log n)$ time~\cite{square}. 
%For each of the valid interval, the upper envelope of the functions  $f_1(*,* ,*)$ can be found in $O(n \log n)$ time. 
Thus, the total running time is $O(n^2\log n)$. 
% Therefore, finding all the valid intervals over all secondary intervals can be done in $O(n^2 \log n)$ time. 
%
%
%Now we will prove that the number of valid intervals is $O(n)$. Within a secondary interval, a blue point will create at most one invalid interval. Therefore, a secondary interval can contain $O(n)$ valid intervals. 
\qed
\end{pf}

	\section{Square Annulus Separability in Fixed Orientation }
	\label{SAS}
	
In this section, we consider the Square Annulus Separability problem. Given a point set $P= \textbf{R} \cup \textbf{B}$, we show how to construct a concentric square annulus of minimum width that separates $\textbf{R}$ and $\textbf{B}$. Let $S$ denote a separating concentric square annulus of minimum width. Let $S_{out}$ and $S_{in}$ respectively be the outer and inner squares that define $S$. The definitions of $\mathcal{R}_{in}$, $\mathcal{B}_{out}$ and fixed points are the same as given in section \ref{RAS}.

%\begin{observation}
%\label{sq1}
%Both $S_{out}$ and $S_{in}$ contain at least one blue point on its boundary.
%\end{observation}
%\textbf{Proof of Observation \ref{sq1}}
%\begin{proof}
%Assume  $S_{out}$ (respectively $S_{in}$) does not contain a blue point on the boundary. Then we can shrink $S_{out}$ (respectively expand $S_{in}$) until it touches a blue point. Then the width of $S$ is reduced. It contradicts that $S$ is of minimum width. 
%\end{proof}

%\begin{proof}
% Assume neither $S_{out}$ nor $S_{in}$ contains a blue point on the boundary. Then we can shrink $S_{out}$ and expand $S_{in}$ till each of them touches a blue point. Then the width of $S$ is reduced. It contradicts that $S$ is minimum width. 
%\end{proof}
%Both $S_{out}$ and $S_{in}$ contain at least one blue point on its boundary.
By reasons similar to those stated in previous sections, it can be assumed that $S$ contains four input points on the boundary. We will refer to them as the \emph{anchor points} of $S$. Equivalently, we can say that $S$ is uniquely \emph{defined by} these four points. Of these, two anchor points are blue points that lie on the boundaries of $S_{out}$ and $S_{in}$.
%In other words, there exists a unique square annulus that contains these four points on the boundary. We say that the square annulus is defined by these four points.
\begin{observation}
\label{sqr2}
$S$ cannot be uniquely defined by four points if one of the following holds.
\newline\noindent 1. Two points are on adjacent edges of $S_{out}$ and the other two points are on adjacent edges of $S_{in}$ and no two of them are on the same side of $S$.
\newline\noindent 2. There is a point each on the left (respectively top) edges of $S_{out}$ and $S_{in}$ and a point each on the right (respectively bottom) edges of $S_{out}$ and $S_{in}$. 
\end{observation}
\noindent Now we can state the following theorem.

\begin{figure}[t]
\begin{minipage}[b]{0.43\linewidth}
\centering
\scalebox{0.25}{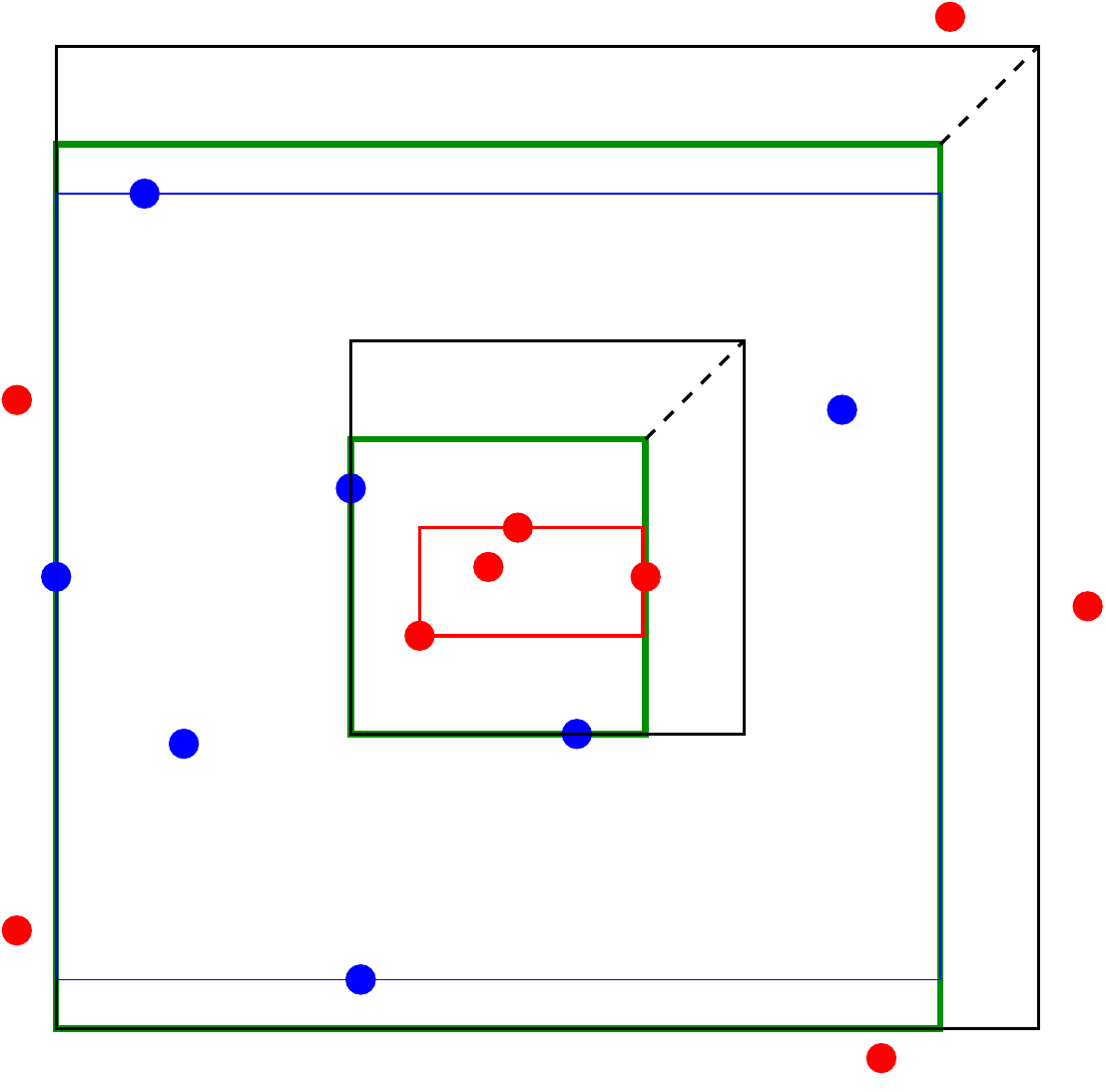}
\caption{The green lines indicate the square annulus after moving through dotted lines.}
\label{fig:figure5}
\end{minipage}
\hspace{0.18cm}
\begin{minipage}[b]{0.48\linewidth}
\centering
\scalebox{0.39}{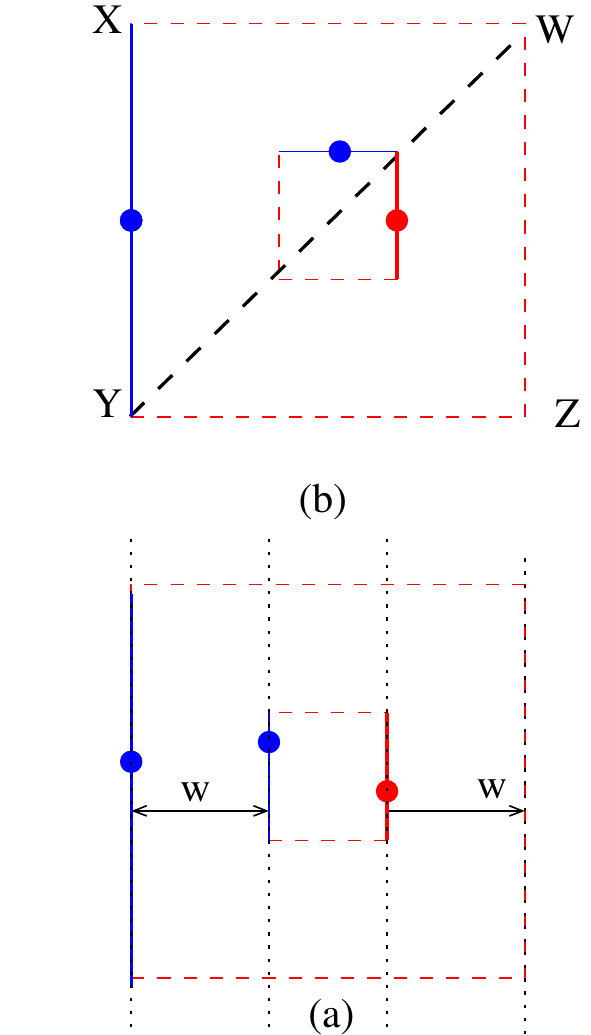}
\caption{(a)Black doted line represents two vertical slabs, (b)Black doted line represents the diagonal of the square.}
\label{fig:figure6}
\end{minipage}
\end{figure}

\begin{thm} 

For a given point set $P= \textbf{R} \cup \textbf{B}$, a concentric square annulus $S$ of minimum width that separates $\textbf{R}$ and $\textbf{B}$ can be constructed in $O(n \log^2 n)$ time.

\end{thm}
\begin{pf}
We start by sorting the blue points and red points by their $x$ and $y$ coordinates. We have seen that four anchor points are enough to define $S$. 

 If there are more than two anchor points which are also fixed points, then we can find $S$ in $O(n \log n)$ time by correctly guessing the anchor points in $O(n)$ time and checking the validity of each of these in $O(\log n)$ time. Therefore assume $S$ contains a fixed point each as anchor points on two opposite sides. If not, there exist two adjacent sides that do not contain any fixed point. Then we can move these two sides inwards along the diagonal until $S_{out}$ touches a point in $\mathcal{B}_{out}$ or $S_{in}$ touches a point in $\mathcal{R}_{in}$  (Fig. \ref{fig:figure5}). Without loss of generality assume that $p_{1}$ and $p_{2}$ are anchor points that lie respectively on the left and right sides of $S$. There should be at least one point from $\mathcal{B}_{out}$. Assume $p_1 \in \mathcal{B}_{out}$. Now there are two cases based on the point $p_2$.

\noindent \textbf{1. The point $p_{2}$ is from $\mathcal{R}_{in}$}:
Now, the third anchor point $p_{3}$ is a blue point in $S_{in}$ since $S_{in}$ always contains a blue point. We can correctly guess this point in $O(n)$ time. There are two cases.\\
\textbf{1.a.} \emph{$p_{3}$ lies on the left edge of $S_{in}$: } Now we know two anchor points $p_{1}$ and $p_{3}$, that lie on the same side of the annulus. This gives us the width of $S$ and thus we know the four vertical lines that contain the left and right edges of $S_{out}$ and $S_{in}$. For a correct guess of anchor points $p_{1}$, $p_{2}$ and $p_{3}$, $R_{in}$ lies between the slabs defined by these lines (Fig. \ref{fig:figure6}(a)). By observation~\ref{sqr2}, we know that the fourth anchor point $p_{4}$ lies on the top or bottom side of $S$. If $p_4$ is a blue point, then it lies on $S_{in}$ and is the nearest blue point vertically from $\mathcal{R}_{in}$. If $p_4$ is a red point, then it lies on $S_{out}$ and is the nearest red point vertically from $B_{out}$. In both the cases $p_{4}$ can be found in $O(\log n)$ time. \\
\textbf{1.b.}\emph{ $p_{3}$ lies on the top edge of $S_{in}$:} By $p_{2}$ and $p_{3}$, we know the top right corner of $S_{in}$ and hence the line containing one diagonal of $S_{in}$. The point where this line intersects the vertical line containing $p_{1}$ is the bottom left corner of $S_{out}$ (Fig. \ref{fig:figure6}(b)). By observation \ref{sqr2}, if $p_{4}$ is a red point, then it lies on the top side or the right side of $S_{out}$. In the former case, it is the red point with minimum $y$-coordinate in the region bounded by the lines $XY$ and $YW$ and in the latter case, it is the red point with minimum $x$-coordinate in the region bounded by the lines $YW$ and $YZ$. If it is a blue point in $S_{in}$, then it is the blue point with the highest $y$-coordinate in the closed polygonal region formed by $YW$, $YZ$ and the vertical line contain $p_{2}$.   \\
The argument is symmetric if the third anchor point that we guess lies on the bottom edge.\\
\noindent\textbf{2. The point $p_{2}$ is from $\mathcal{B}_{out}$} : 
The case where there are two blue anchor points on $S_{in}$ and one of them lies on the left or right edge is similar to 1.a. Otherwise, we construct the rest of the annulus using a slightly different approach. 
Instead of guessing an anchor point,
we use a technique to find the center $c$ of $S$ and the anchor point $p$ that lies on the top or bottom edge of $S_{in}$.
Let $r$ be the radius of $S_{out}$ which is given by half the difference of $x$-coordinates of the left and right edges of $S_{out}$. Let $l$ be the vertical line at distance $r$ from left and right edges of $S_{out}$. The center of the annulus, $c$, should lie on $l$ on a segment $L$ of length at most $2r$. 
We define a function $f_{p}(c)$ for all $p \in \textbf{B}$, $c \in l$ as follows.
%\begin{displaymath}
\begin{equation}
f_{p}(c)=
  \left\{
\begin{array}{ll}
\infty & if \hspace{5pt}  y(t(R_{in})) < y(p) < y(b(R_{in})) \\
\infty & if \hspace{5pt} y(p) > y(t(R_{in}))$ and $y(c)>\\
& (y(p)+(y(b(R_{in}))))/2\\
\infty & if \hspace{5pt} y(p) < y(b(R_{in}))$ and $y(c)< \\
& (y(t(R_{in}))+y(p))/2 \\
|y(p)-y(c)| &  \hspace{10pt}     otherwise
%\end{displaymath}
\end{array}
\right.
\end{equation}

 Intuitively, $f_{p}(c)$ gives the radius of the square centered at $c$ and contains $p$ on the top or bottom edge and contains all red points in $\mathcal{R}_{in}$ inside it. For a fixed $c$, a square centered at $c$ that contains $\mathcal{R}_{in}$ inside and all the blue points outside has radius at most $\min_{p\in \textbf{B}} f_p(c)$. Moreover, a square $S_{in}$ that minimizes the width has radius $max_{c \in l}(\min_{p\in \textbf{B}} f_p(c))$.
 
 It is easy to see that, for a given $p$ the graph of $f_p(c)$ is a straight line segment $L_p$ of slope $+1$ or $-1$. Hence $max_{c \in l}(\min_{p\in \textbf{B}} f_p(c))$ is the maximum value of the lower envelope in the arrangement of $L_p$, for all $p \in \textbf{B}$. Computing the lower envelope of an arrangement of $n$ line segments can be done in $O(n \log n)$ time~\cite{square}.

 Thus, $S$ should have the anchor points in one of a constant number of configurations. In the configuration as in case 1 above, we can correctly guess the anchor points in $O(n \log n)$ time and each guess can be validated in $O(\log n)$ time. Then the total running time for this case is $O(n \log^{2} n)$. A configuration as in case 2 can be solved in $O(n \log n)$ time.
% All other cases can be solved in less time. 
 Thus a separating square annulus of minimum width can be computed in $O(n \log^{2} n)$ time.
 \qed
   \end{pf}

\noindent \textbf{Degenerate cases in Annulus separability}
 
  The  algorithms given in sections \ref{RAS}, \ref{arb} and \ref{SAS} may need slight changes for some degenerate cases. The case where $B_{out}$ does not contain any red point becomes equivalent to the minimum enclosing rectangle/square annulus problem in monochromatic setting and can be solved by the algorithms given in~\cite{joydeep,square}. 
 %In the case of uniform width rectangular annulus, this case is solved by running the algorithm for minimum width non-uniform width rectangular annulus and consider the blue point $b$ that  
 Also, there are cases where one or more sides of the minimum width annulus are at infinity. Such cases can be solved by algorithms that compute separating unbounded L-shapes, separating strips, etc. These can be computed using the same ideas as given above without decreasing the overall efficiency of the algorithm.
 %If there is no $R'$ points then run the algorithm consider $b \in B$ as $R_{in}$ point, where $b$ is from largest width defining side of $A_{in}$. The correctness of this statement is from the fruitful characterization of the minimum width uniform rectangular annulus contain at least one point from $F_p$.

%We can use a similar algorithm to construct a separating concentric square annulus of maximum width.
%\section{Separability using Annulus of Maximum Width}
%We can use the algorithms described above with slight modifications to find the separating annulus of maximum width. For the sake of completeness, we now describe the algorithm to construct the separating concentric square annulus of maximum possible width. Other algorithms in sections \ref{RAS} and \ref{arb} can be modified in a similar way for the corresponding maximum width annulus separability problem.

%---------------------------------------

%	\input{Smax.tex}
	
	\section{Orthogonal Convex Separability}
		\label{OS}

In this section, we give an algorithm to compute the separating orthogonal convex polygon that minimizes the number of edges. 
%\begin{center}
%\begin{figure}
%\scalebox{0.3}{\input{orthogonal.pdf_t}}
%\caption{Orthogonal Convex hull of blue points}
%\end{figure}
%\end{center}

%Orthogonal Convex Hull 
%\footnote{ The orthogonal convex hull of a point set $P$ is defined as the smallest area orthogonal convex polygon that contains $P$. An orthogonal convex hull has its convex vertices as points from $P$}.% The orthogonal convex hull of a point set with $n$ points can be computed in $O(n \log{n})$ time~\cite{Karlsson}.

%For a given point set, let $C$ be a separating orthogonal convex polygon that minimizes the number of edges.

Construct the orthogonal convex hull,
%\footnote{ The orthogonal convex hull of a point set $P$ is defined as the smallest area orthogonal convex polygon that contains $P$. The convex vertices of the orthogonal convex hull of $P$ are points from $P$.},
$\mathcal{H}$, of $\textbf{B}$ in $O(n \log n)$ time~\cite{Karlsson}. 
The orthogonal convex hull of a point set $P$ is defined as the smallest area orthogonal convex polygon that contains $P$. The convex vertices of the orthogonal convex hull of $P$ are points from $P$.
If $\mathcal{H}$ contains red points inside then a separating orthogonal convex polygon does not exist. For a given point $p$, we can check whether $p$ lies inside $\mathcal{H}$ in $O(\log n)$ time. Thus the existence of a separating orthogonal convex polygon can be decided in $O(n \log n)$ time. We now discuss how to construct a separating orthogonal convex polygon, $C$, that minimizes the number of edges.

\begin{figure}[h]
	\centering
	\scalebox{0.23}{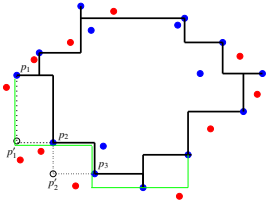}
	\caption{Orthogonal Convex hull of blue points. The green line indicates the orthogonal convex polygon with minimum number of edges.}
	\label{fig:figure1}
\end{figure}
\begin{thm}
	For a given bichromatic point set with $n$ points, a separating orthogonal convex polygon with the minimum number of edges can be computed in $O(n \log n)$ time.
\end{thm}

\begin{pf}

	Let $H$ be the set of vertices of $\mathcal{H}$. Let $p_l$, $ p_b$, $p_r$ and $p_t$ respectively be the leftmost, bottommost, rightmost and topmost points in $H$. Let $P_{lb}$ be the left-bottom chain in $\mathcal{H}$ i.e., the orthogonal path in $\mathcal{H}$ from $p_l$ to $p_b$. We show how to minimize the number of edges in the  left-bottom chain of a separating orthogonal convex polygon. 
	The algorithm is as follows:

	Let $\{p_1,p_2,\dots,p_k\}$ be the convex vertices in $P_{lb}$ taken in counter-clockwise direction. Note that $p_1=p_l$ and $p_k=p_b$. For the purpose of our algorithm, we are only considering the convex vertices of $P_{lb}$.
	% All the convex vertices of $\mathcal{H}$ are points of $B$. 
	Note that $p_1$ and $p_k$ are considered to be convex and the convex vertices can uniquely determine the chain $P_{lb}$. Let $p_i$ be a vertex in this chain, where $1 \leq i < k$ . $p_i$ and $p_{i+1}$ are connected in $\mathcal{H}$ by two axis-parallel line segments such that the horizontal line segment passes through the point $p_i$ and the vertical segment passes through the point $p_{i+1}$.  Let us denote this as the \emph{inward edge} between $p_i$ and $p_{i+1}$. Similarly, let the \emph{outward edge} denote the path comprising of two axis-parallel edges such that the horizontal line segment passes through the point $p_{i+1}$ and the vertical segment passes through the point $p_{i}$. Here, let $p_i^\prime$ denote the point of intersection of the two axis-parallel lines on the outward edge . Let $R_{p_i, p_{i+1}}$ be the rectangle with $p_i$ and $p_{i+1}$ as the diagonal points (see Fig. \ref{fig:figure1}). Starting with $p_1$, we iteratively check every vertex of $P_{lb}$ and see if it can be deleted by replacing the inward edge by the outward edge. Specifically, for all $i$, $1\leq i <k$, we check if the rectangle $R_{p_i, p_{i+1}}$ contains red points inside. If not, we replace the inward edge between $p_i$ and $p_{i+1}$ by the outward edge. In this process, we delete the vertices $p_i$ and $p_{i+1}$ and add the vertex $p_{i}^{\prime}$ to the chain. 
	\vspace{.1cm}
	\noindent\hrule
	\vspace{.1cm}
	\noindent $i=1$, $q_{1}=p_{1}$.\\
	while( $i < k$ )\\
	\indent$q_{2}=p_{i+1}$.\\
	\indent if    $R_{q_{1},q_{2}}$ has a red point, then $q_{1}=q_{2}$.\\
	\indent  else\\
	\indent \indent delete $p_{i},q_{i+1}$ from $C$ and add $p_{i}'$ to $C$, $q_{1}=p_{i}'$\\
	\indent $i=i+1$. end while.\\
	\hrule 
	%\rule{length}{thickness} 

	%If this rectangle contains red points inside, then replace $p_i$ and $p_{i+1}$ respectively by $p_{i+1}$ and $p_{i+2}$ and repeat. Otherwise, if $R_{p_{i}, p_{i+1}}$ does not contain any red point inside, then replace the inward orthogonal edge between $p_i$ and $p_{i+1}$ in $\mathcal{H}$ by the outward axis-parallel edge. Then replace the vertex $p_i$ by $p_{i^\prime{}}$, the intersection point of the vertical line through $p_i$ and horizontal line through $p_{i+1}$ and repeat. Continue this till we reach the point $p_b$
	%Let

	%
	%Our algorithm starts with the orthogonal convex hull which separates red and blue points. It easy to see that there is no red points inside $R_{p_i'',p_{i+1}}$. If there is no red points inside the $R_{p_i,p_{i+1}}$ we can change the inward edge of $\mathcal{H}$ with outward edge of $\mathcal{H}$. Also we change $p_i$ to $p_i'$ in next iteration. 
	%We can do the same procedure until $p_{i+1}$ point become $p_r$. For completing the orthogonal convex object with minimum edge, do the same in  $P_{br}$, $P_{rt}$ and $P_{tl}$ also.
	%\subsection{Algorithm}
	
	\vspace{0.1cm}
	This algorithm returns the left-bottom chain of $C$. To show the correctness, we prove that the left bottom chain of any separating orthogonal convex polygon has at least as many inward edges as returned by our algorithm. Consider a configuration of points $b_1, b_2,r$ such that $b_1, b_2 \in \textbf{B}$, $r \in \textbf{R}$ and $r$ is contained in a rectangle with $b_1$ and $b_2$ as diagonal vertices. For an orthogonal convex polygon to include $b_1$ and $b_2$ and exclude $r$, it should have at least one inward edge. Let $\{q_1,q_2,\dots,q_{k^\prime}\}$ be the set of vertices returned by the algorithm. By the description of the algorithm, every vertex $q_i$ is a blue point or the vertical edge that contains $q_i$ has a blue point above $q_i$. By this property, for an inward edge between $q_i$ and $q_{i+1}$, there is a rectangle $R_i$ such that $R_i$ has two blue points as diagonal points and contains a red point inside. Moreover, for all $i,j$, $1 \leq i,j <k^\prime$, $R_i$ and $R_j$ do not intersect in the interior. This shows that the algorithm returns a chain with the minimum possible number of inward edges possible. We run similar algorithms for all other chains of $\mathcal{H}$- $P_{br}, P_{rt}$ and $P_{tl}$. $C$ is given by the union of these four chains. 
	%The algorithm outputs the separating orthogonal convex polygon with the minimum number of edges. 
	
	%By construction, for every three consecutive vertices in the output polygon, there exists a distinct configuration of three points such that two of them are blue and the other one is red and its x and y coordinates lie between those of the blue points. A separating orthogonal convex polygon that includes the blue points and excludes the red point should have an inward axis parallel edge between the blue points.  
	
	Constructing the orthogonal convex hull  takes $O(n \log n)$ time~\cite{Karlsson}. At every step in the algorithm, we check a rectangle $R_{p,q}$ for red points where $p$ and $q$ are consecutive vertices in the separating polygon constructed so far. After every check, one of these points is deleted from the set of vertices to be considered. Therefore, the number of iterations is linear in the number of points in the orthogonal convex hull. Each check takes $O(\log n)$ time. Thus, the algorithm runs in $O(n\log n)$ time.
	\qed
\end{pf}
By an easy reduction from the sorting problem, this is also optimal.

	\section{Conclusion}
\label{Con}
In this paper, we have studied the geometric separability problem using rectangular annuli of fixed and arbitrary orientation, square annulus of fixed orientation and orthogonal convex polygon. We have designed algorithms for these problems whose running time also matches that of the best known algorithms for the monochromatic variant of these problems. The next interesting questions will be the separability problem for square annulus and orthogonal convex polygon of arbitrary orientation and separability problem using more complex rectilinear objects. Another interesting step is to study the separability problem in higher dimensions.

	
	
	

\end{document}